\documentclass[final]{article}
\usepackage{amscd}
\usepackage{amsxtra}
\usepackage{amssymb, amsmath}
\usepackage{amsthm}

\theoremstyle{plain}
\newtheorem{theorem}{Theorem}[subsection]
\newtheorem{proposition}[theorem]{Proposition}

\newtheorem{definition}[theorem]{Definition}
\newtheorem{remark}[theorem]{Remark}

\begin{document}

\newcommand{\U}{\ensuremath{\mathfrak{u}_{1}}}
\newcommand{\ad}{\ensuremath { ad(\mathfrak{u}_{1})}}
\newcommand{\Oo}{\ensuremath{\varOmega^{0}(ad(\mathfrak{u}_{1}))}}
\newcommand{\Ok}{\ensuremath {\varOmega^{1}(ad(\mathfrak{u}_{1}))}}

\newcommand{\s}{\ensuremath{\mathcal{S}}}
\newcommand{\csa }{\ensuremath {\mathcal{S_{\alpha}}}}
\newcommand{\cs}{\ensuremath{\mathcal{S^{c}}}}
\newcommand{\vsa}{\ensuremath{\varGamma(\mathcal{S^{+}_{\alpha}})}}
\newcommand{\la }{\ensuremath{{\mathcal{L}}_{\alpha}}}
\newcommand{\sla }{\ensuremath{{\mathcal{L}}^{1/2}_{\alpha}}}

\newcommand{\ca}{\ensuremath{\mathcal{C}_{\alpha}}}
\newcommand{\Aa}{\ensuremath {\mathcal{A}_{\alpha}}}
\newcommand{\Q}{\ensuremath{\mathcal{A}_{\alpha}\times_{\mathcal{G}_{\alpha}}
\varGamma (S^{+}_{\alpha})}}
\newcommand{\wQ}{\ensuremath{\mathcal{A}_{\alpha}\times_{\widehat{\mathcal{G}}_{\alpha}}
\varGamma (S^{+}_{\alpha})}}
\newcommand{\B}{\ensuremath{\mathcal{B}_{\alpha}}}
\newcommand{\wB}{\ensuremath{\widehat{\mathcal{B}}_{\alpha}}}

\newcommand{\G}{\ensuremath{\mathcal{G}_{\alpha}}}
\newcommand{\wG}{\ensuremath{\widehat{\mathcal{G}}_{\alpha}}}

\newcommand{\lda}{\lambda}

\newcommand{\spinc}{\ensuremath{Spin^{c}_{4}}}
\newcommand{\spin }{\ensuremath{Spin_{4}}}

\newcommand{\Z }{\ensuremath{\mathbb {Z}}}
\newcommand{\R }{\ensuremath{\mathbb {R}}}
\newcommand{\C }{\ensuremath {\mathbb {C}}}
\newcommand{\N }{\ensuremath{\mathbb {N}}}
\newcommand{\iso }{\ensuremath {\thickapprox }}

\newcommand{\sw}{\ensuremath {\mathcal{S}\mathcal{W}_{\alpha}}}
\newcommand{\y}{\ensuremath {\mathcal{Y}\mathcal{M}}}
\newcommand{\yp}{\ensuremath {\mathcal{Y}\mathcal{M}^{+}}}
\newcommand{\yn}{\ensuremath {\mathcal{Y}\mathcal{M}^{-}}}
\newcommand{\Cf }{\ensuremath {\mathcal{C}_{\alpha}}}
\newcommand{\w }{\ensuremath {\omega}}
\newcommand{\cx}{\ensuremath {C_{X}}}

\title{A  Necessary Condition for the Existence of $\sw$-Monopoles}
\author{ Celso M. Doria \\ UFSC - Depto. de Matem\'atica }
\maketitle
\begin{abstract}

\noindent Originally, as described in ~\cite{Wi94}, the $\sw$-equations discovered by Seiberg 
and Witten  are $1^{st}$-order partial differential equations, which solutions $(A,\phi)$, with $\phi\ne 0$,
 are known as $\sw$-monopoles. It is known that the solutions of these $1^{st}$-order $\sw$-equations 
correspond to the minimum of the functional $\sw:\ca\rightarrow\R$. However, it is not true that the minimum of 
$\sw:\ca\rightarrow\R$ is always attained by this sort of solution. 
In fact, there are only a finite number of $\alpha\in Spin^{c}(X)$ such that the minimum is a
$\sw$-monopole. We show that a necessary condition to $(A,\phi)\in \ca$  be a $\sw$-monopole is
 that   

$$\frac{Q_{X}(\alpha,\alpha)}{v_{X}.(k^{-}_{g,X})^{4}}\in [-\frac{1}{\pi^{2}},\frac{1}{4}],$$

\noindent  where $Q_{X}$=intersection form of $X$, $v_{X}$=volume of $(X,g)$ and 
$k^{-}_{g,X}$ is a constant depending on the 
scalar curvature of $(X,g)$.
 
 \end{abstract}

\section{Introduction}

Let $(X,g)$ be a closed Riemannian 4-manifold and $\alpha\in Spin^{c}(X)$ be a $spin^{c}$-structure
fixed on X. The space of $Spin^{c}$-strutures on X is 

 $$Spin^{c}(X)=\{\alpha\in H^{2}(X,\mathbb{Z})\mid w_{2}(X) = \alpha \ mod
 \ 2\}.$$  

For each $\alpha \in Spin^{c}(X)$,  there is a vector spave $V$ and  a representation 
$\rho_{\alpha}:SO_{4}\rightarrow Gl(V)$. Consequently, there are also a pair of vector bundles 
$(\csa^{+},\la)$ over X (see ~\cite{LM89}), where ($P_{SO_{4}}$=frame bundle of $TX$)
\begin{itemize}
\item $\csa = P_{SO_{4}}\times_{\rho_{\alpha}} V = \csa^{+}\oplus
  \csa^{-}$.\\
 The bundle  $\csa^{+}$ is the positive complex spinors bundle (fibers are 
$Spin^{c}_{4}-modules$ isomorphic to $\C^{2}$) 
\item $\la=P_{SO_{4}}\times_{det(\alpha)} \C$.\\
 It is called the \emph{determinant line bundle} associated to the
 $Spin^{c}$-struture $\alpha$. ($c_{1}(\la)=\alpha$)
\end{itemize}

Thus, given $\alpha \in Spin^{c}(X)$ we associate a pair of bundles

$$\alpha \in Spin^{c}(X) \quad \rightsquigarrow \qquad(\la,\csa^{+})$$

Let $P_{\alpha}$ be the frame bundle of $\la$. So, $c_{1}(P_{\alpha})=\alpha$.

Now, consider the spaces
\begin{itemize}
\item $\mathcal{A}_{\alpha}= L^{1,2}(\Oo)$ 
\item $\varGamma(\csa^{+})$ = $L^{1,2}(\varOmega^{0}(X,\csa^{+})$  
\item $\ca = \mathcal{A}_{\alpha}\times \varGamma(\csa^{+})$
\item $\G= L^{2,2}(X,U_{1})=L^{2,2}(Map(X,U_{1}))$
\end{itemize}

In dimension 4, the vector bundle $\varOmega^{2}(ad(\mathfrak{u}_{1}))$ admits a decomposition

\begin{equation}
\varOmega^{2}_{+}(ad(\mathfrak{u}_{1}))\oplus \varOmega^{2}_{-}(ad(\mathfrak{u}_{1}))
\end{equation}

\noindent in  seld-dual (+) and anti-self-dual (-) parts (~\cite{DK91}).

\vspace{05pt}

 The 1$^{st}$-order (original) \emph{Seiberg-Witten} equations are defined  over the configuration space
$\ca =\mathcal{A}_{\alpha}\times \varGamma(\csa^{+})$ as

\begin{equation}\label{E:03}
\begin{cases}
D^{+}_{A}(\phi )= 0,\\ 
F^{+}_{A} = \sigma (\phi)
\end{cases}
\end{equation}  
where
\begin{itemize}
\item $D^{+}_{A}$ is the $Spinc^{c}$-Dirac operator defined on $\vsa$;
\item $\sigma:\varGamma(\csa^{+})\rightarrow End^{0}(\vsa^{+})$ ($End^{0}(V)$=traceless endomorphism of V)
is the  quadratic form 

\begin{equation}\label{E:QF}
\sigma (\phi)=\phi\otimes\phi^{*} - \frac{\mid\phi\mid^{2}}{2}.I
\end{equation}

\vspace{05pt}

\noindent performing the coupling of the \emph{ASD}-equation with
the $Dirac^{c}$ operator. \\
Locally, if $\phi = (\phi_{1},\phi_{2})$, then the quadratic form
$\sigma(\phi)$ is written as 

\begin{equation*}
 \sigma(\phi) = \left(
 \begin{matrix}
 \frac{\mid\phi_{1}\mid^{2}-\mid\phi_{2}\mid^{2}}{2} & \phi_{1}.\Bar{\phi_{2}} \\
 \phi_{2}.\Bar{\phi_{1}} &\frac{ \mid\phi_{2}\mid^{2} - \mid\phi_{1}\mid^{2}}{2}
 \end{matrix}
 \right)
 \end{equation*}
\end{itemize}

\section{\bf{A Variational Principle for the Seiberg-Witten Equation}}

Consider the functional

\begin{equation}\label{E:SW00}
SW (A,\phi) =\frac{1}{2} \int_{X}\{\mid F^{+}_{A} - \sigma(\phi)\mid^{2} + \mid
D^{+}_{A}(\phi)\mid^{2}\}dv_{g}
\end{equation}

\noindent The next identities, which proofs are standard, are applied to 
expand the functional (~\ref{E:SW00})

\begin{proposition}
For each $\alpha \in Spin^{c}(X)$, let $\mathcal{L}_{\alpha}$ be the
determinant line  bundle associated to $\alpha$ and (A,$\phi$) $\in\ca$. Also, assume that 
$k_{g}$=scalar curvature of (X,g). Then,

\begin{enumerate}
\item $<F^{+}_{A},\sigma(\phi)> = \frac{1}{2} <F^{+}_{A}.\phi ,\phi >$
\vspace{04pt}
\item $<\sigma(\phi ),\sigma (\phi )> =\frac{1}{4}\mid \phi \mid^{4} $
\vspace{04pt}
\item Weitzenb$\ddot{o}$ck formula $$ D^{2}\phi = \triangledown^{*}\triangledown \phi +
  \frac{k_{g}}{4}\phi +\frac{F_{A}}{2}.\phi $$

\item $\sigma(\phi)\phi = \frac{\mid\phi\mid^{2}}{2}\phi$
\vspace{04pt}
\item $c_{2}(\la\oplus \la)=\int_{X}F_{A}\wedge F_{A} $
\vspace{04pt}
\item $\mid F^{+}_{A}\mid^{2} = \frac{1}{2}\mid F_{A}\mid^{2} - 4\pi^{2}\alpha^{2}$
\end{enumerate}
\end{proposition}

Consequently, after expanding the functional (~\ref{E:SW00}), we  get the expression

\begin{equation}\label{E:SW01}
SW(A,\phi) = \int_{X}\{\frac{1}{4}\mid F_{A}\mid^{2} + \mid \triangledown^{A} \phi
\mid^{2} + \frac{1}{8}\mid \phi\mid^{4} + \frac{1}{4}<k_{g}\phi , \phi> \}dv_{g}- 2\pi^{2}\alpha^{2}
\end{equation}

\vspace{05pt} 

\begin{definition}
For each $\alpha\in Spin^{c}(X)$, the Seiberg-Witten Functional is the functional 
 $\sw:\ca\rightarrow\R$ given by
 \begin{equation}\label{E:SW02}
\sw(A,\phi) = \int_{X}\{\frac{1}{4}\mid F_{A}\mid^{2} + \mid \triangledown^{A} \phi
\mid^{2} + \frac{1}{8}\mid \phi\mid^{4} + \frac{1}{4}<k_{g}\phi , \phi> \}dv_{g}
\end{equation}
where $k_{g}$= scalar curvature of (X,g).
\end{definition}

Let $k^{m}_{g,X}=\min_{x\in X}k_{g}(x)$ and 

\begin{equation}\label{E:SC}
k^{-}_{g,X}=
\begin{cases}
0,\quad \text{if $k_{g}(x)\ge 0$ for all $x\in X$};\\
(- k^{m}_{g,X})^{\frac{1}{2}}, \quad\text{otherwise}.
\end{cases}
\end{equation}

\begin{remark}
  \begin{enumerate}
\item Since X is compact and $\mid\mid \phi \mid\mid_{L^{4}}<\mid\mid \phi
    \mid\mid_{L^{1,2}}$, the functional is well defined on $\ca$ ,
\item The $\sw$-functional \emph{(\ref{E:SW02})} is Gauge invariant.

\end{enumerate}
\end{remark}

\begin{proposition}
The Euler-Lagrange equations of the $\sw$-functional \emph{(\ref{E:SW02})} are

\begin{equation}\label{E:09}
\Delta_{A} \phi + \frac{\mid \phi \mid^{2}}{4}\phi +\frac{k_{g}}{4}\phi = 0 
\end{equation}
\begin{equation}\label{E:10}
d^{*}F_{A} + 4\Phi^{*}(\triangledown^{A}\phi)=0
\end{equation}
where $\Phi:\varOmega^{1}(\mathfrak{u}_{1})\rightarrow\varOmega^{1}(\csa^{+})$
\end{proposition}

\begin{remark}
\begin{enumerate}
\item From (~\ref{E:SW00}) and (~\ref{E:SW01}), it follows that 

$$\sw (A,\phi)-2\pi^{2}\alpha^{2}\ge 0.$$

\noindent The equality above happens iff $(A,\phi)$ is a solution to the $1^{st}$-order $\sw$-equations.
Thus, the $\sw$-functional is bounded below by $2\pi^{2}\alpha^{2}$, where 
$$\alpha^{2}= Q_{X}(\alpha , \alpha)$$
 ($Q_{X}:H^{2}(X,\Z)\times H^{2}(X,\Z)\rightarrow \Z$ is the
 intersection form of X). 
\vspace{04pt}
\item It is known from (~\cite{JPW96}) that whenever $(A,\phi)$ is a solution of (~\ref{E:09}) and 
(~\ref{E:10}), then

\begin{equation}\label{E:12}
\mid\mid\phi\mid\mid_{\infty}\le k^{-}_{g,X}
\end{equation}
\end{enumerate}
\end{remark}

It follows from  Holder's Inequality that 

$$\int\mid\phi\mid^{2}\le [\int
dv_{X}]^{1/2}.[\int\mid\phi\mid^{4}]^{1/2} \Rightarrow
[ \int\mid\phi\mid^{2}]^{1/2}\le v^{1/4}_{X}.[\int\mid\phi\mid^{4}]^{1/4},$$

\begin{equation}\label{E:11}
\mid\mid\phi\mid\mid_{L^{2}}\le v^{1/4}_{X}\mid\mid\phi\mid\mid_{L^{4}}
\end{equation} 

\noindent where $v_{X}$ is the volume of X.

The next proposition was orinally proved  in ~\cite{Wi94}.
 
\begin{proposition}
Let $\alpha\in Spin^{c}(X)$. Let $(A,\phi)$ be a $\sw$-monopole, so 

\begin{equation}\label{E:L}
\alpha^{2}\le \frac{v_{X}.(k^{-}_{g,X})^{4}}{4}
\end{equation}

\end{proposition}
\begin{proof}
It follows fro the Weitzenb$\ddot{o}$ck formula that

$$D_{A}\phi=0\quad\Rightarrow\quad (\triangledown^{A})^{*}\triangledown^{A}\phi + 
\frac{1}{2}F^{+}_{A}(\phi) + \frac{1}{4}k_{g}\phi=0$$

\noindent By taking the inner product with $\phi$ and integrating, we get

$$\int_{X}\left(\mid\triangledown^{A}\phi\mid^{2} + \frac{1}{4}\mid\phi\mid^{4} +
\frac{1}{4}k_{g}\mid\phi\mid^{2}\right)dx=0$$

\noindent Consequently, applying (~\ref{E:12}), we get

$$\int_{X}\left(\mid\triangledown^{A}\phi\mid^{2} + \frac{1}{4}\mid\phi\mid^{4}\right)dx \le 
 \frac{1}{4}(-k^{m}_{g,X})\int_{X}\mid\phi\mid^{2}dx\le \frac{(k^{-}_{g,X})^{4}.v_{X}}{4}.$$

\noindent Since $\phi\ne 0$, we can apply the last inequality to estimate 
$\mid\mid F^{+}_{A}\mid\mid_{L^{2}}$;

$$\mid\mid F^{+}_{A}\mid\mid^{2}_{L^{2}}=\mid\mid\sigma(\phi\mid\mid^{2}_{L^{2}} = 
\frac{1}{4}\mid\mid\phi\mid\mid^{4}_{L^{4}}\le  \frac{(k^{-}_{g,X})^{4}.v_{X}}{4}$$

\noindent Once $\alpha^{2}=\mid\mid F^{+}_{A}\mid\mid^{2}_{L^{2}}-\mid\mid F^{-}_{A}\mid\mid^{2}_{L^{2}}$, it 
follows that

$$\alpha^{2}\le \frac{(k^{-}_{g,X})^{4}.v_{X}}{4}.$$

\end{proof}

\begin{proposition}
Let $\alpha\in Spin^{c}(X)$ and   $(A,\phi)$ be a $\sw$-monopole. So, 

\begin{equation}\label{E:L}
\alpha^{2}\ge -\frac{v_{X}.(k^{-}_{g,X})^{4}}{\pi^{2}}.
\end{equation}

\end{proposition}
\begin{proof}

$$\sw(A,\phi) - \int_{X}k_{g}\mid\phi\mid^{2} = \frac{1}{4}\mid\mid F_{A}\mid\mid^{2}_{L^{2}} + 
\mid\mid\triangledown^{A}\phi\mid\mid^{2}_{L^{2}} + \frac{1}{8}\mid\mid\phi\mid\mid^{4}_{L^{4}}  $$

\noindent Therefore,

$$\sw(A,\phi) + (-k^{m}_{g,X})\mid\mid\phi\mid\mid^{2}_{L^{2}}\ge \frac{1}{4}\mid\mid F_{A}\mid\mid^{2}_{L^{2}} + 
\mid\mid\triangledown^{A}\phi\mid\mid^{2}_{L^{2}} + \frac{1}{8}\mid\mid\phi\mid\mid^{4}_{L^{4}},$$

\noindent in particular,

$$\mid\mid\phi\mid\mid^{4}_{L^{4}} \le 8\sw(A,\phi) + 8(k^{-}_{g,X})^{2}\mid\mid\phi\mid\mid^{2}_{L^{2}}$$

\noindent Applying the inequality (~\ref{E:11}), we get that

$$\mid\mid\phi\mid\mid^{4}_{L^{2}} \le 8v_{X}\sw(A,\phi) + 8v_{X}.(k^{-}_{g,X})^{2}\mid\mid\phi\mid\mid^{2}_{L^{2}}.$$ 

\noindent So,

\begin{equation}\label{E:12}
\mid\mid\phi\mid\mid^{4}_{L^{2}} - 8v_{X}.(k^{-}_{g,X})^{2}\mid\mid\phi\mid\mid^{2}_{L^{2}}  - 8v_{X}\sw(A,\phi) \le 0.
\end{equation}

\noindent Consider the quadratic function $f:\R\rightarrow\R$ defined as 

$$f(x)= x^{2} - 8v_{X}.(k^{-}_{g,X})^{2}x  - 8v_{X}\sw(A,\phi)$$

\noindent If the inequality $f(x)\le 0$  is not satisfied by any $x\in\R$, then 
$\phi=0$. The discriminant  of f is

$$\triangle = 32v_{X}\left( 2v_{X}.(k^{-}_{g,X})^{4}_{g,X} + \sw(A,\phi)\right)$$

\noindent The inequality (~\ref{E:12}) will be satisfied if and only if $\triangle \ge 0$, since 
$f^{``}(x)>0$ and $\min_{x\in\R}f(x)=-\frac{\triangle}{4}$. Therefore,

$$\sw(A,\phi)\ge -2v_{X}.(k^{-}_{g,X})^{4}$$

\noindent It follows that

$$4v_{X}.(k^{-}_{g,X})^{2} - 2\sqrt{4v_{X}\left[2v_{X}(k^{-}_{g,X})^{4} + \sw(A,\phi)\right]} 
\le \mid\mid\phi\mid\mid^{2}_{L^{2}} \le $$

$$\le 4v_{X}.(k^{-}_{g,X})^{2} + 2\sqrt{4v_{X}\left[2v_{X}.(k^{-}_{g,X})^{4} + \sw(A,\phi)\right]}$$

\noindent Since $\sw(A,\phi)\ge 2\pi^{2}\alpha^{2}$, the lower upper bound of the $\sw$-functional is

$$\sw(A,\phi)\ge \max\{2\pi^{2}\alpha^{2}, - 2v_{X}.(k^{-}_{g,X})^{4}\}$$

\noindent In this way,  if $(A,\phi)$ ($\phi\ne 0$) satisfies the  $1^{st}$-order $\sw$-equation then

$$\alpha^{2}\ge -\frac{v_{X}.(k^{-}_{g,X})^{4}}{\pi^{2}}$$

\end{proof}

The inequalities obtained impose necessary concitions in order to $(A,\phi)$ be a $\sw$-monopolo. 
Basically, it is necessary that

\begin{equation}\label{E:20}
-\frac{v_{X}.(k^{-}_{g,X})^{4}}{\pi^{2}}\le \alpha^{2} \le \frac{(k^{-}_{g,X})^{2}.v_{X}}{4}
\end{equation} 

\noindent or, equivalently, 

$$\frac{\alpha^{2}}{v_{X}.(k^{-}_{g,X})^{4}}\in [-\frac{1}{\pi^{2}},\frac{1}{4}]$$

\vspace{20pt}

\noindent\emph{Universidade Federal de Santa Catarina \\
    Campus Universitario , Trindade\\
               Florianopolis - SC , Brasil\\
               CEP: 88.040-900}
\vspace{10pt}

\end{document}